\documentclass[aps,pre,nofootinbib,reprint,superscriptaddress]{revtex4-2} 

\usepackage[utf8]{inputenc}
\usepackage{graphicx}
\usepackage{subfigure}
\usepackage{xcolor}
\usepackage{subfigure}
\usepackage{placeins}

\usepackage{xcolor}

\usepackage{enumitem}

\newcommand{\be}{\begin{eqnarray}}
\newcommand{\ee}{\end{eqnarray}}

\newcommand{\wbe}{\begin{widetext}}
\newcommand{\wee}{\end{widetext}}

\newcommand{\eq}[1]{(\ref{#1})}

\usepackage{bbm}
\usepackage{amsmath,leftidx}
\usepackage{graphicx}
\usepackage{times}
\usepackage{CJK}
\usepackage{color,colortbl}
\usepackage[normalem]{ulem}

\usepackage[
colorlinks=true,        
citecolor=blue,         
linkcolor=blue,         
urlcolor=blue           
]{hyperref}             

\begin{document}

\title{ Subsystem Thermalization and Work Statistical Characterizations of Floquet Dynamics}

\author{Feng-Li Lin}
\email{fengli.lin@gmail.com}
\affiliation{Department of Physics, \\
National Taiwan Normal University, Taipei, 11677, Taiwan}

\author{Ching-Yu Huang}
\email{cyhuangphy@thu.edu.tw}
\affiliation{Department of Applied Physics, \\
Tunghai University, Taichung 40704, Taiwan}
\affiliation{ Physics Division, National Center for Theoretical Sciences, Taipei 10617, Taiwan}

\begin{abstract}

We study Floquet thermalization in a periodically driven quantum non-integrable Ising chain by combining two operational diagnostics: subsystem thermalization and work statistics. For generic interacting Floquet systems, stroboscopic dynamics lead to heating toward an infinite‑temperature state at low driving frequencies, to a prethermal state during the crossover regime, and then to a finite-temperature state at high driving frequencies. We show that reduced density matrices of small subsystems provide a precise measure of local equilibration and clearly resolve prethermal plateaus. In parallel, we analyze the statistics of work performed over a Floquet cycle using both the characteristic function of work with and without the two-point measurements, and the related fluctuation theorem, which captures coherent contributions and deviations from thermal equilibrium. 
By comparing these two diagnostics within the same Floquet setting, we demonstrate that work statistics encode the same dynamical crossover that governs subsystem thermalization. Our results establish a unified and experimentally accessible framework for characterizing Floquet thermalization, prethermal regimes, and coherent energy absorption in interacting quantum systems.
\end{abstract}

\date{\today}

\maketitle


\section{Introduction}

Periodically driven quantum many-body systems provide a controlled setting for exploring nonequilibrium statistical mechanics. A central theme is \emph{Floquet thermalization} , the tendency of generic interacting systems to absorb energy from the drive and approach an infinite-temperature state. Foundational works \cite{Lazarides2014FloquetETH, DAlessio2014FloquetThermalization, Ponte2015FloquetErgodic, Kuwahara2016FloquetMagnus, Abanin2017EffectiveHamiltonians} established that Floquet eigenstates typically satisfy a version of the eigenstate thermalization hypothesis (ETH) \cite{Srednicki_1999, Srednicki:1995pt, Srednicki1994ETH, Deutsch1991ETH, DAlessio:2015qtq, deutsch2018eigenstate} adapted to periodically driven systems. High-frequency expansions and Floquet--Magnus techniques clarified how approximate integrals of motion can stabilize long-lived prethermal regimes before eventual heating~\cite{Bukov2015AdvancesFloquet, Mori:2016mtn, Else2016TimeCrystalsPrethermal, Abanin2017PrethermalizationRigorous, Weidinger2017SciPostPrethermal}. In parallel, studies of many-body localization (MBL) demonstrated that disorder can suppress heating entirely, stabilizing nonthermal Floquet phases~\cite{Ponte2015FloquetMBL, Lazarides2015MBLProtection, Abanin2019RMPMBL, Khemani2016FloquetMBLPhases}. Together, these results provide a detailed picture of how Floquet systems interpolate between prethermal plateaus, constrained dynamics, and the universal infinite-temperature steady state.

A complementary line of research focuses on \emph{subsystem thermalization} \cite{Dymarsky:2016ntg, Lashkari:2016vgj, He:2017vyf, He:2017txy, Lin:2024vji}, which examines whether reduced density matrices of small subsystems approach Gibbs or generalized Gibbs ensembles determined by the relevant conserved quantities. This perspective is closely tied to ETH, typicality, and entanglement growth, and has been used to characterize equilibration in static, quenched, and driven non-Hermitian systems~\cite{Rigol2008NatureETH, Linden2009SubsystemThermalization, GogolinEisert2016Review, DAlessio2016ETHReview, lin2024work}. Subsystem thermalization provides an operational and experimentally accessible probe of local equilibration scales, emergent conservation laws, and the distinction between genuinely thermal behavior and constrained or quasi-integrable dynamics.

A second complementary viewpoint arises from quantum thermodynamics, where the energetic cost of driving is characterized through \emph{work statistics}. In closed quantum systems, work is defined operationally, most commonly via the two-point measurement (TPM) protocol, which yields a classical probability distribution of work, and the fluctuation theorem for the thermal state \cite{Jarzynski1997Equality, Crooks1999Fluctuation, Kurchan2000QuantumFluctuation, Talkner2007WorkNotObservable, Campisi2011RMPFluctuation, Esposito2009RMP}.  The fluctuation theorem is the modern formulation of the second law of thermodynamics in quantum thermodynamics and can be used to detect global thermality. The latter contrasts with the detection of local thermalization by the subsystem thermalization protocol. Beyond TPM, full counting statistics (FCS) \cite{nazarov2003full, tobiska2005inelastic, esposito2009nonequilibrium, Solinas:2015wne} and the associated characteristic functions retain quantum coherence and allow one to define \emph{quasi-probability distributions} of work, which can exhibit negativity and encode time-reversal asymmetry ~\cite{Mukamel2003QuantumJarzynski, Solinas2016QuasiProbWork, PerarnauLlobet2017NoGoWork, Perarnau-Llobet:2017qee, Alonso2016TimeAsymmetryWork, Funo2018PathIntegralWork, Lostaglio:2018ogq, Zhang:2022msk, Levy:2020sii, Gherardini:2024ije}. These tools underpin quantum fluctuation relations and provide a bridge between microscopic unitary evolution and thermodynamic notions such as irreversibility and entropy production.

In this work, we bring these two operational diagnostics---subsystem thermalization and work statistics---together in the context of Floquet thermalization. We consider interacting periodically driven systems and analyze, within the same setting, (i) the reduced density matrices of subsystems to be compared with their counterparts of target thermal states, and (ii) the statistics of work performed over one or multiple Floquet cycles, including TPM fluctuation theorem, quasi-probabilistic characteristic functions, and the related coherence quantities. Subsystem thermalization provides a local probe of equilibration and prethermal plateaus, while work statistics quantify global energy absorption and coherent contributions to the drive-induced dynamics. By comparing their behavior across parameter regimes, we show that both diagnostics capture the same dynamical crossover from prethermal regimes to eventual heating, and that signatures of emergent thermality and coherent freezing appear consistently in both the reduced states and the work distribution.

Our results thus establish a unified framework in which Floquet thermalization is characterized simultaneously by local subsystem properties and global work statistics. This framework provides a more systematic characterization of Floquet thermalization than earlier studies, based on recent developments in quantum thermodynamics, such as ETH/subsystem thermalization, work statistics/fluctuation theorem, and extensions to quasi-probabilities. This aims to provide experimentally accessible tools for diagnosing prethermalization, heating, and the breakdown of coherent dynamics in Floquet quantum matter.

The remainder of this paper is organized as follows. In the next section, we outline the framework for characterizing Floquet thermalization using indicators based on subsystem thermalization and work statistics, including coherence and fluctuation-theorem-related quantities. In section \ref{number}, we present numerical results for indicators of Floquet thermalization, based on which we examine the consistency between subsystem thermalization and work statistics, and discuss their implications. Finally, we conclude our paper in section \ref{conclude}. In Appendix \ref{appA} we provide the details on deriving the quantum work average from the Schrodinger equation, and in Appendix \ref{appB} we present the results of a finite-scaling study on the coherence-related quantities.

\section{Framework}

In this section, we will outline our framework for characterizing the Floquet thermalization. 

\subsection{Floquet Dynamics}

Floquet dynamics is a cyclic driving process of a closed quantum system under unitary evolution. The setup is similar to that for the fluctuation theorem of work statistics under a cyclic process, except that the initial state changes from cycle to cycle. Thus, the Floquet thermalization can be viewed from the perspective of the generalized fluctuation theorem and work statistics.

\begin{figure}[bth]
\includegraphics[width=0.45\textwidth]{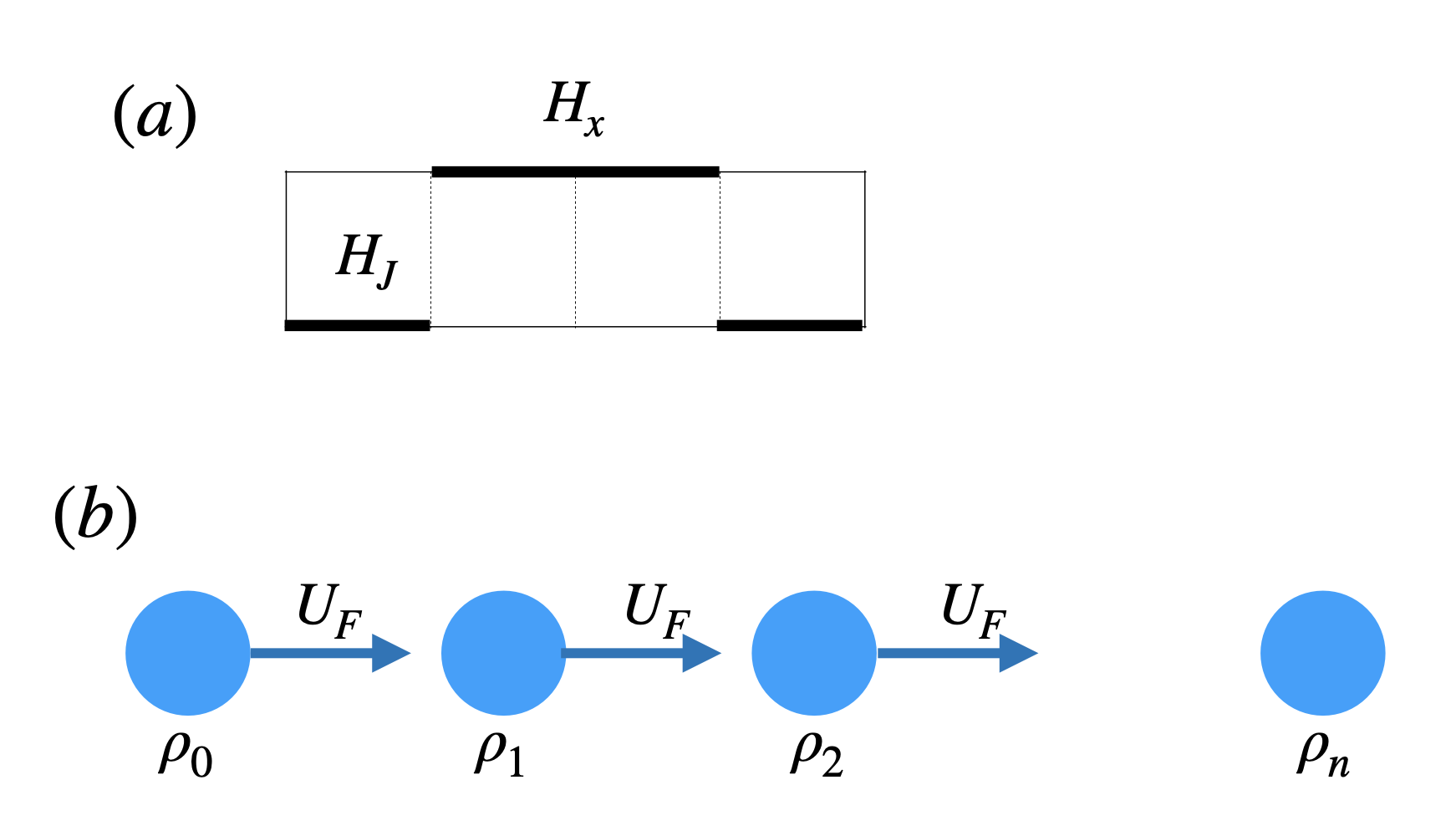}
\caption{\small  
(a) Schematic illustration of the periodically driven Hamiltonian. Within one driving period T, the system alternates between two non-commuting parts,  $H_J$ and  $H_x$, in a piecewise-constant (bang–bang) manner.
(b) Stroboscopic evolution of the quantum state under the Floquet operator  $U_F$. Starting from the initial state $\rho_0$, the state after $n$ driving cycles is given by $\rho_n$, where each application of $U_F$  corresponds to one full driving period $2\pi/\omega$. }
\label{Floquet} 
\end{figure}

In this work, we study the Floquet dynamics of an $N$-site open Ising spin chain described by the following Hamiltonian
\be\label{H}
H=H_J + \lambda(t) H_x
\ee
with $\lambda(t)$ the cyclic driving profile of characteristic frequency $\omega$, and 
\be\label{HJ}
H_J &=& -J \sum_{i=1}^N \sigma_i^z \sigma_{i+1}^z - h_z \sum_i \sigma_i^z\;, 
\\ \label{Hx}
H_x &=& -h_x \sum_{i=1}^N \sigma_i^x\;,
\ee
where $\sigma_i^{x,y,z}$ are the Pauli matrices of the $i$-th site spin. Thus, Floquet dynamics is dictated by the so-called Floquet operator
\be
U_F := e^{-i \int_0^{2\pi/\omega} dt H(t) dt} = e^{- i \frac{2\pi}{\omega} H_{\rm eff}},
\ee
where we have introduced the effective Hamiltonian $H_{\rm eff}$ encoding the nontrivial Dyson series. Given an initial state $\rho_0$ of the Floquet process, one can then trace the final state of the $n$-th Floquet cycle, denoted by $\rho_n$, i.e.,
\be
\rho_n=\big[ U_F \big]^n \rho_0  \big[ U_F^{\dagger} \big]^n \;, \quad n=1,2,\cdots
\ee
which is also the final state of the $(n+1)$-th Floquet cycle. See Fig. \ref{Floquet}(b).

From the above setup, it is straightforward to postulate that the thermalization of the Floquet dynamics shall depend on
\begin{enumerate}[label=(\roman*)]
\item  the characteristic frequency $\omega$\;;

\item the initial state $\rho_0$ of the whole Floquet process\;;

\item  the target thermal state at the $n$-th Floquet cycle,
\be\label{target_thermal}
\rho_{\beta_n} := {e^{-\beta_n H_0}  \over {\rm Tr} e^{-\beta_n H_0}}
\ee
for some choice of $H_0$ and its conjugate inverse temperature $\beta_n$. If Floquet thermalization occurs, the system's final state $\rho_n$ at the $n$-th Floquet cycle should approach this thermal state as $n\gg 1$.

\end{enumerate}

Each choice corresponds to a different physical notion of thermalization.

We now elaborate on each item in the above. 
First, the previous studies \cite{Mori:2016mtn} show that the Floquet system can be thermalized to $\beta_{n\gg 1}=0$ thermal state for small $\omega < \log N$ \cite{DAlessio:2013ugl, DAlessio:2014rzv, lazarides2014equilibrium, Kim:2014jfl,  ponte2015periodically}, and to finite $\beta_{n \gg 1}$ thermal state for large $\omega \gtrsim N$, and to so-called Floquet prethermalization \cite{Else:2016ags, abanin2017effective, mori2018floquet, Ho:2022ene} for $\omega$ in between. 

Second, we can implement Floquet dynamics for two categories of $\rho_0$: (a) the canonical thermal states at some temperature, and (b) the pure states, such as the Néel state or the ground state of $H_J$. As the Floquet dynamics are unitary, a pure initial state will remain pure under the process, so the Floquet system cannot be globally thermalized but can be locally thermalized. This motivates the present work to examine Floquet thermalization via subsystem thermalization, which will depend on $\omega$, as expected. On the other hand, an initial thermal state may evolve into another thermal state under many cycles of Floquet dynamics with either small or large values of $\omega$. In this case, Floquet thermalization can be examined through subsystem thermalization and the passivity of thermal states, as derived from the fluctuation theorem.

{Third, a central issue in characterizing Floquet thermalization is the choice of the reference Hamiltonian $H_0$ used to define the target thermal state $\rho_{\beta_n}$. 
Unlike static systems, periodically driven systems naturally admit several nonequivalent Hamiltonian descriptions, and therefore no unique prescription exists for selecting the thermal ensemble. 
Once $H_0$ is specified, the effective inverse temperature $\beta_n$ can be determined from the energy-matching condition}
\be\label{be_beta_n}
{\rm Tr}[\rho_n H_0 ] = {\rm Tr}[\rho_{\beta_n} H_0]\;. 
\ee
In this work, we consider three interesting options:
\begin{enumerate}
\item $H_0 = H_J$\;,

\item $H_0 = H_{\rm eff}$\;,

\item  $H_0= H_{\rm ave} := {1\over 2}(H_J + H_x)$ \;. 
\end{enumerate}
Options 1 and 2 are more intuitive, and the last option is due to our choice of driving profile, as shown in Fig. \ref{Floquet}(a), by taking the average period of $H_J$ and $H_x$. By the Floquet-Magnus expansion \cite{Viebahn2020IntroductionTF}, $H_{\rm eff}$ and $H_{\rm ave}$ are different by the higher order non-commutative terms, e.g.,
\be\label{F_M_exp}
H_{\rm eff} = H_{\rm ave} + {i \pi \over 4 \omega} [H_J,H_x] + {\cal O}\big(\omega^{-2}\big)\;.
\ee
It is clear that $H_{\rm eff} \simeq H_{\rm ave}$ only if $\omega \gg 1$ or $[H_J, H_x]\simeq 0$, which, however, is not the case in the current consideration.  

Guided by the above discussions on the postulate for Floquet thermalization, we will study the patterns of subsystem thermalization and the statistical Characterization of work by varying $\omega$, $\rho_0$, and $H_0$ to examine whether and how Floquet thermalization emerges.  
Since thermal states are passive and satisfy the fluctuation theorem, deviations from these properties of work statistics offer an operational probe of the thermalization process.  

 The implication of this target thermal state is twofold. If $\rho_0$ is a pure state, the unitary Floquet evolution will just turn it into another pure state, which, however, can be a typical state so that the target thermal state \eq{target_thermal} can be adopted to characterize the subsystem thermalization. We then refer to this kind of Floquet thermalization as "local (Floquet) thermalization." 
 
On the other hand, if the initial state $\rho_0$ is a thermal state, then the Floquet dynamics could drive it to another thermal state, which we postulate to be \eq{target_thermal} and shall obey the fluctuation theorem in the context of work statistics. We call this kind of  Floquet thermalization the "global (Floquet) thermalization." 

Below, we elaborate on how to use subsystem thermalization and work statistics to characterize Floquet thermalization before presenting the numerical results.

\subsection{Subsystem Thermalization Characterization}

The eigenstate thermalization hypothesis (ETH) postulates that a given energy eigenstate appears locally thermal. The subsystem thermalization then extends ETH to generic many-body entangled states, or the so-called typical states, which are expected to be locally thermal as well. It can be examined by calculating the relative entropy of the reduced density matrix of the given state $\rho$ and the corresponding thermal state $\rho_{\beta}$ in a local region $A$ (with its complement denoted by $\bar{A}$), which are denoted by respectively $\rho^A:= {\rm Tr}_{\bar{A}}\rho$ and $\rho^A_{\beta}={\rm Tr}_{\bar{A}}\rho_{\beta}$,
\be
S\big[ \rho^A \big\Vert  \rho^A_{\beta} \big] := {\rm Tr}\big[ \rho^A \big( \ln \rho^A -\ln \rho^A_{\beta} \big)\big]\;,
\ee
where the inverse temperature of the thermal state $\rho_{\beta}$ is determined by requiring
\be\label{be_beta}
{\rm Tr}[\rho H] = {\rm Tr}[\rho_{\beta} H]
\ee
where $H$ is the Hamiltonian of the system, so that $\rho_{\beta}={e^{-\beta H} \over {\rm Tr}[e^{-\beta H}}$. Thus, if the subsystem thermalization holds, then for small enough $A$ one shall have \cite{Mueller:2013bww, Lin:2024vji}
\be\label{subsys_th}
S\big[ \rho^A \big\Vert  \rho^A_{\beta} \big] \simeq 0\;.
\ee

Although the ETH or subsystem thermalization is usually considered for $\rho$ to be a pure state, i.e., the energy eigenstates or typical states, one can extend to the mixed state by using the same criterion given by \eq{subsys_th}. Based on this, we can then apply the subsystem thermalization to examine whether the thermalization of Floquet state $\rho_n$ by checking if 
\be\label{subsys_th_n}
S\big[ \rho^A_n \big\Vert  \rho^A_{\beta_n} \big] \simeq 0\;,
\ee
where $\beta_n$ is determined by \eq{be_beta_n}, which is the Floquet version of \eq{be_beta}. Again, the choice of $H_0$ is essential to determine $\beta_n$ and to examine \eq{subsys_th_n}. From our numerical study, we will see that choosing $H_0 = H_{\rm ave}$ yields more sensible results for Floquet thermalization.

If \eq{subsys_th_n} holds up to $A=N/2$, then it  provides strong evidence of global thermalization. 
Otherwise, the Floquet thermalization can be at least local thermalization. If $\rho_0$ is a pure state, then we expect $\rho_n$ to be thermalized by Floquet dynamics only locally, i.e., $A \ll N$. On the other hand, if $\rho_0$ is a thermal state with inverse temperature $\beta_0$, then the Floquet state $\rho_n$ can be globally thermalized. We will see that our numerical results agree with the above expectation.

\subsection{Coherence Characterization by work statistics}

The work $W$ done by a driving force during a specific period, such as in a Floquet cycle, is not a definite observable but a random variable. Thus, one should consider the work statistics characterized by the work distribution function (WDF) $P(W)$ (with $\int_{-\infty}^{\infty} dW P(W)=1$), which can be constructed by the so-called two-point measurement (TPM) scheme. This was done by performing the energy measurement of the initial and final state of each Floquet cycle in the energy eigenbasis,
\be
 H_J :=\sum_k \varepsilon_k \Pi_k
\ee
with 
with $\varepsilon_k$'s the eigenenergies of $H_J$ and $\Pi_k$'s the corresponding projectors. Note that  $\Pi_k \Pi_l = \Pi_k \delta_{k,l}$ so that $[H_J,\Pi_k]=0$, and $\sum_k \Pi_k = \mathrm{I}$. 
The TPM protocol employs the projective energy measurements, which will dephase (i.e., destroy the coherence of) an observable ${\cal O}$ (which can also be state $\rho$) in the energy eigenbasis, i.e., taking the  dephasing average in the energy eigenbasis,  
\be\label{dephasing_1}
{\cal O} \; \Longrightarrow \; \overline{\cal O} \equiv \sum_k \Pi_k {\cal O } \Pi_k\;.
\ee

The transition probability of the TPM scheme for the $n$-th Floquet cycle is then given by
\be
p_n(k,k')=\mathrm{Tr}\big[\Pi_{k'} U_F \Pi_k \rho_{n-1}\Pi_k U_F^\dagger \big]\;.
\ee
and the associated WDF can be constructed by
\be\label{PW_TPM}
P^{\rm TPM}_n(W) = \sum_{k, k'} p_n(k,k')\, \delta(W-\varepsilon_{k'}+\varepsilon_k)\;.  
\ee
Fourier transform it to get the associated characteristic function
\be
\chi_n^{\rm TPM}(u) &\equiv& \int_{-\infty}^{\infty} dW e^{i u W} P^{\rm TPM}_n(W)\;, 
\\
&=& \sum_{k, k'} p_n(k, k') e^{i u(\epsilon_{k'}-\epsilon_k)}\;, \\ \label{chi_TPM}
&=&  {\rm Tr}\Big[e^{i H_J u} U_F e^{-i u H_J} \overline{\rho_{n-1}} U^{\dagger}_F \Big]\;,
\ee
which depends only on the incoherent part of the initial state.

We have used the dephasing notation introduced in \eq{dephasing_1} for $\overline{\rho_{n-1}}$, which can only be reduced to $\rho_{n-1}$ if it is diagonal in the energy eigenbasis. From the above characteristic function, we can obtain the average work done during the $n$-th Floquet cycle, 
\be\label{W_ave_TPM}
\langle W \rangle^{\rm TPM}_n &=& -i \lim_{u\rightarrow 0} \partial_u \chi_n^{\rm TPM}(u) 
\\
&=& {\rm Tr}\big[H_J ( U_F \overline{\rho_{n-1}} U^{\dagger}_F -  \rho_{n-1} ) \big]\;,
\ee
where $\langle \cdots \rangle^{\rm TPM}_n \equiv \int_{-\infty}^{\infty} d W P^{\rm TPM}_n(W) \cdots $ To arrive at the second equality, we have used the fact ${\rm Tr}[H_J \overline{\rho_{n-1}}] ={\rm Tr}[H_J \rho_{n-1}]$. 

On the other hand, we can identify the quantum work gain in the time duration dt to be
\be
\delta W= {d H(t) \over dt} dt = {d H \over d\lambda} {d\lambda \over dt} dt = H_x {d\lambda \over dt} dt \;,
\ee
which can be considered as a quantum statistical observable due to the involvement of $d\lambda/dt$, which yields statistical uncertainty in the context of work statistics.
Then, using the Schrodinger equation 
\be
{d\rho(t)\over dt}=-i[H(t),\rho(t)]
\ee
for the system's state $\rho(t)$, we can obtain the average work done as follows  (see Appendix \ref{appA} for details),
\be\label{Mech_W}
\langle W \rangle_n  = \int_{(n-1)T}^{n T} {\rm Tr}\big[ \delta W \rho(t) \big] = {\rm Tr}[H_J (\rho_n - \rho_{n-1})]\;. \qquad 
\ee
Note that $\langle W \rangle_n$ is generally different from  $\langle W \rangle^{\rm TPM}_n$ because $\rho_n = U_F \rho_{n-1} U^{\dagger} \ne U_F \overline{\rho_{n-1}} U^{\dagger}_F$ if $\rho_{n-1}$ is not not diagonal in the energy egigenbasis. 

Since the TPM destroys the quantum coherence, it is motivated to define the coherence-preserving characteristic function (of $n$-th Floquet cycle) as follows:
\be\label{chi_FCS}
\chi_n(u)={\rm Tr}[ e^{i H_J u} U_F e^{-i H_J u} \rho_{n-1} U^{\dagger}_F]\;.
\ee
It is the generalization of full counting statistics (FCS) \cite{nazarov2003full, tobiska2005inelastic, esposito2009nonequilibrium, Solinas:2015wne}, which measures the fluctuation of the quantity $\Sigma = B(T) - A(0)$ during a cyclic process with initial state $\rho$:  
\be
\chi(u) = {\rm Tr}[e^{i B(T) u} U(T) e^{-i A(0) u} \rho U^{\dagger}(T) ]\;,
\ee
which can be treated as a physical observable by measuring it through Ramsey-like interferometry \cite{Dorner:2013mes, Mazzola:2013gmy, Mazzola:2014dvw}. However, due to the quantum coherence, the inverse transform of $\chi_{\Sigma}(u)$ will generally give a quasiprobability 
\be
P(\Sigma) = {1\over 2\pi}\int_{-\infty}^{\infty} du \; e^{-i u \Sigma} \chi(u)\;, 
\ee
unless the initial state $\rho$ is an incoherent state in the energy eigenbasis. 

\bigskip

We can apply the above basic summary of work statistics to characterize Floquet thermalization. 
As a thermal state should be a maximally incoherent state  in the eigenbasis of $H_J$, 
the quantum coherence of a Floquet state can be used to tell the degree of its thermalization.
Following the key observation that the TPM destroys the quantum coherence of the Floquet states, we can characterize it in two ways: 

\begin{enumerate}

\item If the state $\rho_{n-1}$ is not incoherent, it can be detected by the quantity
\be
C_n^{W} &:=& {1\over N} \big[ \langle W \rangle^{\rm TPM}_n - \langle W \rangle_n \big] \;, \\
&=& {1\over N} {\rm Tr}[H_J U_F (\overline{\rho_{n-1}} -\rho_{n-1}) U_F^{\dagger}]\;,
\ee
which characterizes how the average work per site done by the driving force is affected by the coherence of the initial state.

\item Since the quantum coherence will be destroyed by TPM, yielding $\chi^{\rm TPM}_n(u)$ to differ from $\chi_n(u)$, so that the quantum coherence of $\rho_{n-1}$ can also be characterized by the coherent part of $\chi_n(u)$, denoted by
\be
\chi_n^{\rm coh}(u)=\chi_n(u)-\chi_n^{\rm TPM}(u)\;.
\ee
Thus, the deviation from the Floquet thermalization at the $n$-th cycle can be measured by 
\be
C_n^{\chi}=\int_{-\infty}^{\infty} du \; \vert \chi_n^{\rm coh}(u) \vert\;.
\ee
If $\chi_n^{\rm coh}(u)$ is periodic in $u$ with period $T_u$, then replace $\int_{-\infty}^{\infty} du$ in the above by ${1\over T_u}\int_{T_u} du$.
\end{enumerate}

Therefore, both $C_n^W$  and $C_n^{\chi}$ quantify the \sout{residual} coherence of the Floquet state  calibrated by TPM in the $H_J$-eigenbasis. If the Floquet dynamics drive the system toward a thermal state associated with $H_J$, both quantities are expected to decay toward zero as $n$ increases.

\bigskip

\subsection{Fluctuation-theorem characterization}

Among the incoherent states, the thermal states are peculiar as their work statistics satisfy the fluctuation theorem, i.e., Jarzynski relation, for the cyclic processes \cite{jarzynski1997nonequilibrium, crooks1999entropy, crooks2000path}:
\be\label{FT_chi}
\chi^{\rm TPM}(i\beta)=\langle e^{-\beta W}\rangle^{\rm TPM} =1 \;.
\ee
where $\beta$ is the inverse temperature of the initial thermal state. By Jensen's inequality, this relation implies the second law $\langle W \rangle^{\rm TPM} \ge 0$ for exciting a thermal state. 

Motivated by the Jarzynski relation \eq{FT_chi} for thermal states, naively, we can estimate the deviation of the Floquet state from the thermal state by measuring the following quantity
\be
\big\vert \chi^{\rm TPM}_{n+1}({i\beta_n}) -1 \big\vert
\ee
with $\beta_n$ determined by \eq{be_beta_n} by choosing $H_0=H_J$. However, the fluctuation theorem can, in fact, be generalized to the Kubo-Martin-Schwinger (KMS) states \cite{Pusz:1977hb, Lenard1978ThermodynamicalPO}, among which the thermal state is a special case. 
{A KMS state $\rho_{\rm KMS}$ is associated with a modular operator $\Delta_{\rm mod}$ (or the associated modular Hamiltonian $H_{\rm mod} = -\ln \Delta_{\rm mod}$) so that it satisfies the KMS condition 
\be
{\rm Tr}[\rho_{\rm KMS} A B] = {\rm Tr}[\rho_{\rm KMS} B \Delta_{\rm mod} A]\;,
\ee
which then implies that $\rho_{\rm KMS} \propto e^{-\beta H_{\rm mod}}$ for some parameter $\beta$. When $H_{\rm mod}$ is the physical Hamiltonian, then $\rho_{\rm KMS}$ is a thermal state. }
We can then define a more general characteristic function  during the $(n+1)$-th Floquet by 
\be
\chi^{{\rm TPM}, H_{\rm mod}}_{n+1}(u) =  {\rm Tr}\Big[e^{i H_{\rm mod} u} U_F e^{-i H_{\rm mod} u} \overline{\rho_n} U^{\dagger}_F \Big]\;, \quad
\ee
{for which $H_{\rm mod}$ can be chosen to be $H_J$ or $H_{\rm ave}$ for the current case to check if the (modular) fluctuation theorem holds, i.e., if $\chi^{{\rm TPM}, H_{\rm mod}}_{n+1}(i \beta_n) = 1?$ If the fluctuation theorem holds approximately, it implies $\rho_{n}\propto e^{-\beta H_{\rm mod}}$, which is either a thermal state for $H_{\rm mod} =H_J$ or a KMS state for $H_{\rm ave}$. Note that  $H_{\rm mod} =H_{\rm eff}$ is excluded, because $[H_{\rm eff}, U_F]=0$ and $\chi^{{\rm TPM}, H_{\rm mod}=H_{\rm eff}}_{n+1}(u)=1$ holds trivially. 
Therefore, we can characterize the deviation of $\rho_n$ from a KMS state by the quantity
\be\label{CnFT_def}
C_n^{\rm FT}=\big\vert \chi^{{\rm TPM}, H_{\rm mod}}_{n+1}(i \beta_n) - 1\big\vert
\ee
with $H_{\rm mod} = H_J$ or $H_{\rm ave}$ with $\beta_n$ determined by \eq{be_beta_n} with $H_0=H_{\rm mod}$. Therefore, $C_n^{\rm FT}$ quantifies the extent to which the Floquet state violates the fluctuation theorem associated with the reference modular Hamiltonian $H_{\rm mod}$. A vanishing value indicates consistency with a KMS description, whereas a finite value signals a departure from it.
}

Up to now, we have considered two different ways of characterizing the Floquet thermalization. The first is the subsystem thermalization characterization by the relative entropy $S\big[ \rho^A \big\Vert  \rho^A_{\beta} \big]$. Such thermalization should occur for a typical state for small subsystem size, i.e., $A\ll N$, and the late-time Floquet state should be typical, as expected for the entanglement spreading by long-time nontrivial evolution.
 The bipartite relative entropy for a pure state reaches its maximum at equal bipartition, i.e., $A=N/2$. If this maximal relative entropy is tiny, it is quite plausible that the corresponding Floquet state is well approximated by the thermal state $\rho_{\beta_n}$, i.e., implying global Floquet thermalization. We can examine this speculation by considering the corresponding $C_n^{\rm FT}$ or its coarse-grained version, which probes deviations from a global thermal/KMS state. Given that global Floquet thermalization is well approximated by the same target thermal state, we expect the characterizations by subsystem thermalization and the fluctuation theorem to agree in such cases.

\begin{figure*}[bth]
\includegraphics[width= 1.0\textwidth]
{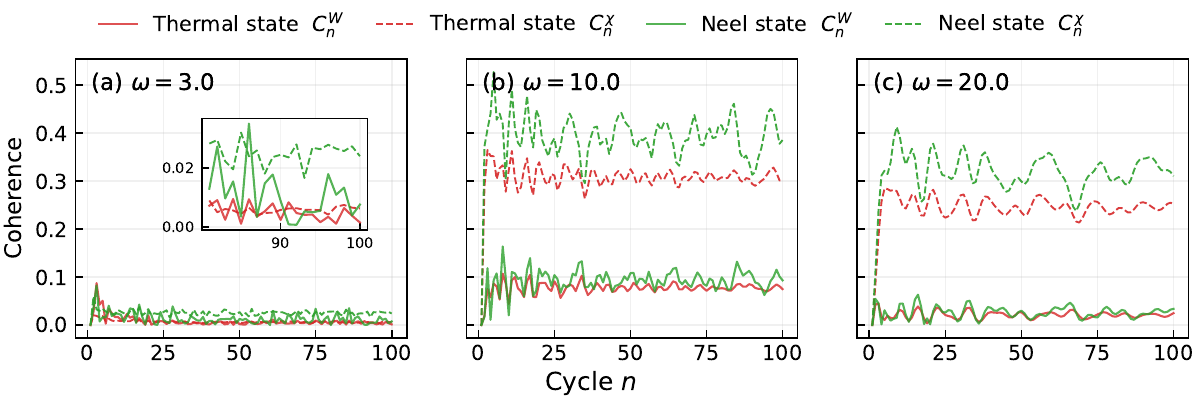}
\caption{\small 
Stroboscopic evolution of the coherence-related quantities $C_n^W$ (solid lines) and $C_n^{\chi}$ (dashed lines) for characterizing the quantum coherence of Floquet dynamics by work statistics from the initial state $\rho_0$ which is either a thermal state $\rho_{\beta=1}$ (red) or a Néel state (green). for three driving frequencies, $\omega=3.0, 10.0, 20.0$. The inset figure in the subfigure (a) shows the zoom-in pattern in the regime $80 \le n \le 100$. For this figure and the ones presented later, the initial states are always prepared with respect to the spin chain Hamiltonian $H_J + H_x$ of $N=10$ sites with fixed parameters $J=1$, $h_x=0.7$, and $h_z=1.0$. 
}
\label{thermal_coherence}
\end{figure*} 

\section{ Numerical characteristics of Floquet thermalization}\label{number}

We now present the numerical results by the aforementioned ways for the characterization of Floquet thermalization. As the coherence characterization requires no determination of $\beta_n$ and no choice of $H_0$, we will present these results first, followed by those from subsystem thermalization and the fluctuation theorem. All the numerical results in this section are for the $N=10$-site spin chain Hamiltonian given in \eq{HJ} and \eq{Hx} with fixed parameters $J=1$, $h_x=0.7$, and $h_z=1.0$. 

Subsystem thermalization will not depend strongly on system size, and the fluctuation theorem and work statistics hold for all system sizes. Thus, we will consider an $N=10$-site spin chain, so that all the stroboscopic evolution under Floquet dynamics presented below can be exactly carried out with reasonable computing resources by the standard techniques of exact diagonalization. 
To assess finite-size effects, we additionally perform a finite-size scaling analysis of the coherence quantities in Appendix~\ref{appB}.

\subsection{Characterization by quantum coherence}

In Fig. \ref{thermal_coherence}, we show the stroboscopic evolution of the coherence-related  quantities $C_n^W$ (solid lines) and $C_n^{\chi}$ (dashed lines) from an initial state $\rho_0$ to be either $\beta_{\beta=1}$ (red) or a Neel state (green), for three driving frequencies $\omega=3.0, 10.0, 20.0$ with the initial state to be the thermal state $\rho_{\beta=1}$. If Floquet thermalization occurs, we expect these coherence quantities to approach a small value as the thermal states are highly incoherent. 

 For the cases of $\rho_0=\rho_{\beta=1}$, i.e., without initial quantum coherence, we see that the coherence quantities $C_n^W$ and $C_n^{\chi}$ rise at the early stage of their stroboscopic evolution, and then soon settle down and fluctuate around some plateau values. The plateau value and the magnitude of fluctuation depend on the driving frequency. We will not compare the sizes of $C_n^W$ and $C_n^{\chi}$ as they are defined and normalized differently. However, it is sensible to compare the sizes of the same quantity at different driving frequencies. 

At low frequency ($\omega=3.0$), both coherence quantities approach zero with tiny fluctuation for large enough $n$. It reflects that the final Floquet state is $\rho_{\beta=0}$, implying that the huge thermal fluctuations destroy the quantum coherence completely.

On the other hand, at high frequency ($\omega=20.0$), both coherence quantities approach a respective value higher than the $\omega=3.0$ counterpart, with also sizable fluctuations. Finally, at intermediate frequency ($\omega=1.0$), both coherence quantities approach a respective value slightly higher than the $\omega=20.0$, but with a similar size of fluctuations. 
It implies that the Floquet states at high driving frequencies are less coherent than those at intermediate driving frequencies. 
The persistence of finite coherence at $\omega=10.0$ and $\omega=20.0$ suggests incomplete thermalization and is consistent with Floquet prethermal behavior, whereas low-frequency driving leads to efficient thermalization and loss of coherence.

 For the cases of $\rho_0$ to be a Neel state, we see that the stroboscopic evolutions of the coherence quantities follow the same patterns as for the cases $\rho_0=\rho_{\beta=1}$, except that their generic sizes are larger than their $\rho_{\beta=1}$ counterparts. 
 This is expected because the Neel states possess initial quantum coherences that could remain manifest under unitary Floquet evolution, except at low driving frequency.  
 This implies that the dephasing of a pure state arises from low-frequency cyclic driving, which can yield global scrambling of the system state over large length scales. 

\begin{figure*}[bth]
\includegraphics[width=1.0\textwidth]{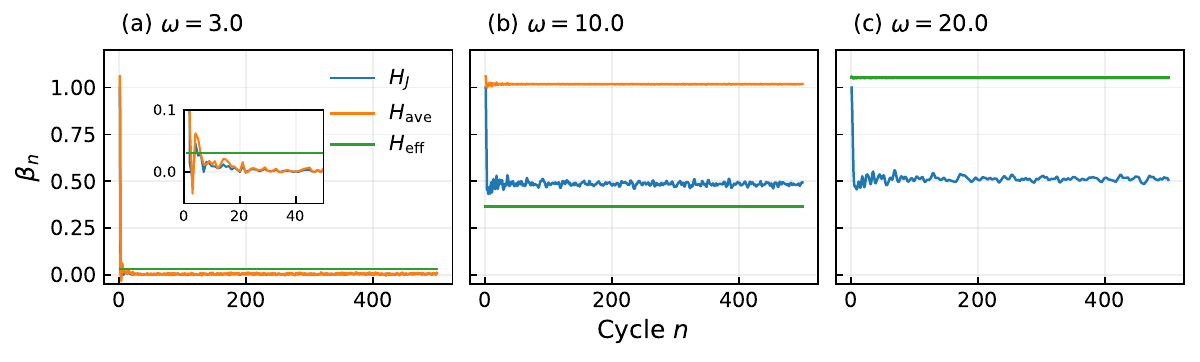}
\caption{\small Stroboscopic evolution of the effective inverse temperature $\beta_n$ defined by \eq{be_beta_n} by choosing $H_0= H_J$ (blue), $H_{\rm ave}$ (orange) and $H_{\rm eff}$ (green) with three driving frequencies $\omega=3.0, 10.0, 20.0$ as shown in the respective subfigures (a), (b), (c) from an initial thermal state of unit inverse temperature, $\rho_0=\rho_{\beta=1}$. The result is briefly summarized as follows. In the subfigure (a), $\beta_{n\gg 1}$ approaches zero for $H_J$ and $H_{\rm ave}$, but not for $H_{\rm eff}$ as shown in the zoomed inset figure. In the subfigure (b), the $\beta_{n\gg 1}$'s for the three Hamiltonians are different. In the subfigure (c), $H_{\rm ave}$ and $H_{\rm eff}$ yield the same $\beta_{n\gg 1}$, which is slightly larger than $\beta_0=1$ but is larger than the $\beta_{n\gg 1}$ from $H_J$.  
 }
\label{thermal_beta} 
\end{figure*} 

\subsection{Characterization by subsystem thermalization}

As discussed, we can examine Floquet thermalization more directly by considering subsystem thermalization and deviation from the fluctuation theorem. The former concerns local thermalization, and the latter concerns global thermalization. For both considerations, we need to postulate an ansatz thermal state whose inverse temperature $\beta_n$ is determined by \eq{be_beta_n} with an appropriate choice of $H_0$. Thus, we first need to study the stroboscopic evolution of the effective inverse temperature $\beta_n$ for three choices of $H_0$, i.e., $H_J$, $H_{\rm ave}$, and $H_{\rm eff}$, and then decide a better option. 

\subsubsection{For initial thermal state}

The stroboscopic evolution under Floquet dynamics can start with either a thermal state or a Néel state. 
Since a pure state remains pure under unitary evolution, global thermalization cannot occur without coarse-graining, 
so we will consider the evolution from a thermal state with unit inverse temperature ($\rho_{\beta=1}$),
which may evolve toward a state that can be approximated by a canonical ensemble under Floquet dynamics.
The results for considering three driving frequencies $\omega=3.0, 10.0, 20.0$ are shown in Fig. \ref{thermal_beta}.

At low frequency ($\omega=3.0$), both $H_J$ and $H_{\rm ave}$ yield $\beta_{n\gg 1}=0$ but $H_{\rm eff}$ yields small but nonzero value.
Thus, our result suggests that the Floquet state approaches the infinite-temperature state at low driving frequency.
At the intermediate frequency $\omega=10.0$, where a Floquet prethermal regime is expected, the asymptotic inverse temperatures obtained from the three choices of $H_0$ differ significantly.
Comparing the intermediate ($\omega=10.0$) and high-frequency ($\omega=20.0$) cases, the asymptotic values obtained from $H_J$ and $H_{\rm ave}$ remain nearly unchanged, whereas the value associated with $H_{\rm eff}$ exhibits a noticeable difference.
Furthermore, $H_{\rm ave}$ and $H_{\rm eff}$ yield the same $\beta_{n \gg 1}$ at high frequency.
This agreement is expected in the high-frequency limit, where the Floquet-Magnus expansion~\eq{F_M_exp} implies $H_{\rm eff}\simeq H_{\rm ave}$,
and with the expectation that the Floquet states approach a thermal state with nonzero $\beta$. 

Based on Fig. \ref{thermal_beta}, we find that the canonical ensemble associated with $H_{\rm ave}$ provides the most physically reasonable description of the effective temperature over the entire frequency range considered. In particular, it correctly reproduces the approach to the infinite-temperature state at low frequency and remains consistent with the high-frequency Floquet-Magnus limit. Further evidence supporting this choice will be provided by the subsystem thermalization and fluctuation-theorem analyses presented below. 

\begin{figure*}[bth]
\includegraphics[width= 1.0\textwidth]{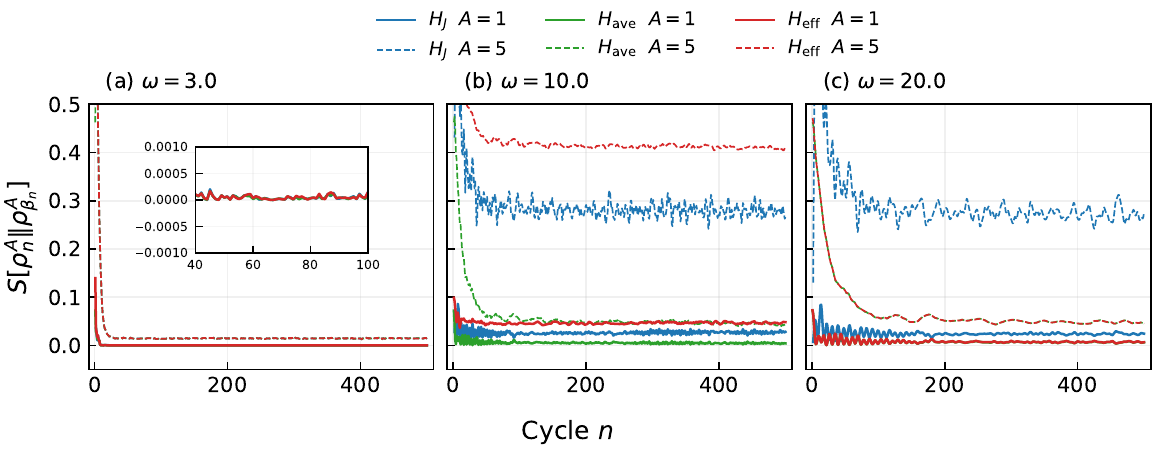}
\caption{\small 
Comparison of stroboscopic behaviors of relative entropy $S\big[ \rho^A_n \big\Vert  \rho^A_{\beta_n} \big]$ in characterizing subsystem Floquet thermalization for the $\beta_n$ given in Fig. \ref{thermal_beta} for $H_0=H_J$ (blue), $H_{\rm ave}$ (green) and $H_{\rm eff}$ (red) from a initial thermal state $\rho_0=\rho_{\beta=1}$. We again consider three driving frequencies (a) $(\omega=3.0)$, (b) $(\omega=10.0)$, and (c) $(\omega=20.0)$ with two different subsystem sizes $A=1$ (solid lines) and $A=5$ (dashed lines). At low frequency ($\omega = 3.0$), the relative entropy approaches almost vanishing values for both $A=1$ and $A=5$ for all three choices of $H_0$. It implies that the subsystem Floquet thermalization works well up to $A=5$. At the intermediate and high frequencies ($\omega =10.0, 20.0$), $H_{\rm ave}$ always yields better subsystem thermalization behaviors for both $A=1$ and $A=5$ than $H_J$ and $H_{\rm eff}$. Moreover, $H_J$ yields quite fluctuating results for $A=5$ even at large $n$. These results imply that $H_{\rm ave}$ will be the better option when considering the thermalization characteristics involving $\beta_n$.
}
\label{thermal_SA_Hall}
\end{figure*} 

Using the inverse temperature $\beta_n$ obtained in Fig. \ref{thermal_beta}, we next investigate subsystem thermalization through the relative entropy
$S\big[\rho_n^A\Vert \rho_{\beta_n}^A\big]$. To maintain consistency with the previous analysis, we first consider the initial thermal state $\rho_{\beta=1}$. The results for subsystem sizes $A\!=\!1$ and $A\!=\!5$ (half of the spin chain) are shown in Fig. \ref{thermal_SA_Hall}.
As already shown in Fig. \ref{thermal_beta}, the large $n$ Floquet states approach an infinite-temperature thermal state at low frequency ($\omega =3.0$); we therefore expect the late-time relative entropy to be almost zero. 
As expected,  all three choices of Hamiltonian give almost the same stroboscopic evolutions by approaching zero relative entropy for $A\!=\!1$ and a tiny one for $A\!=\!5$.
This universal behavior is consistent with the fact that the infinite-temperature state is maximally mixed, making the reduced density matrices nearly indistinguishable from those of the corresponding thermal states.
Since $A\!=\!5$ corresponds to the half-chain subsystem, the tiny relative entropy indicates that thermalization extends over a substantial fraction of the system. A more stringent test of global thermalization will be provided below through the fluctuation-theorem analysis.

As we move into the intermediate ($\omega = 10.0$) and high ($\omega = 20.0$) frequencies, the above universal features disappear. Despite that, $H_{\rm ave}$ still yields almost zero relative entropy for $A=1$ at large $n$ for both frequencies. 
{ A large-$n$ Floquet state is possibly a typical state, as the cyclic unitary evolutions provide long-time scrambling of the quantum information encoded in the initial state. Subsystem thermalization should hold for typical states and manifest as a vanishing relative entropy, at least for small $A$. Thus, the fact that $H_{\rm ave}$ yields the smallest large-$n$ relative entropy indicates that it provides a more appropriate ansatz thermal state.} 
Moreover, at $\omega=20.0$, we again see that the stroboscopic patterns from $H_{\rm ave}$ and $H_{\rm eff}$ coincide for both $A=1$ and $A=5$. 
This agreement is expected in the high-frequency limit, where the Floquet-Magnus expansion predicts $H_{\rm eff}\simeq H_{\rm ave}$.
On the other hand, for $A=5$, $H_J$ yields considerably larger and more strongly fluctuating relative entropy at both frequencies, and $H_{\rm eff}$ yields a larger large-$n$ relative entropy at intermediate frequency.

Combining the results of Figs.~\ref{thermal_beta} and \ref{thermal_SA_Hall} and the related discussions for considering the Floquet thermalization starting from the thermal state $\rho_{\beta=1}$, we conclude that: The most appropriate  ansatz canonical ensemble thermal state $\rho_{\beta_n}$ for determining $\beta_n$ by \eq{be_beta_n} is 
\be\label{ansatz_thermal}
\rho_{\beta_n} = {e^{-\beta_n H_{\rm ave}} \over {\rm Tr} e^{-\beta_n H_{\rm ave}}}\;.
\ee
Adopting this ansatz state, the relative entropy of the large-$n$ Floquet state vanishes at all frequencies considered if $A=1$, but only at low frequency if $A=5$. 
{These results indicate that subsystem thermalization persists over increasingly large spatial scales only in the low-frequency regime. 
{Once the subsystem thermalization holds up to half the system's size, the underlying quantum state could be globally thermalized, or well approximated by a thermal state.} Whether global Floquet thermalization is achieved {in such a regime} will be examined further through the fluctuation-theorem analysis presented below.}

\begin{figure*}[bth]
\includegraphics[width= 1.0\textwidth]{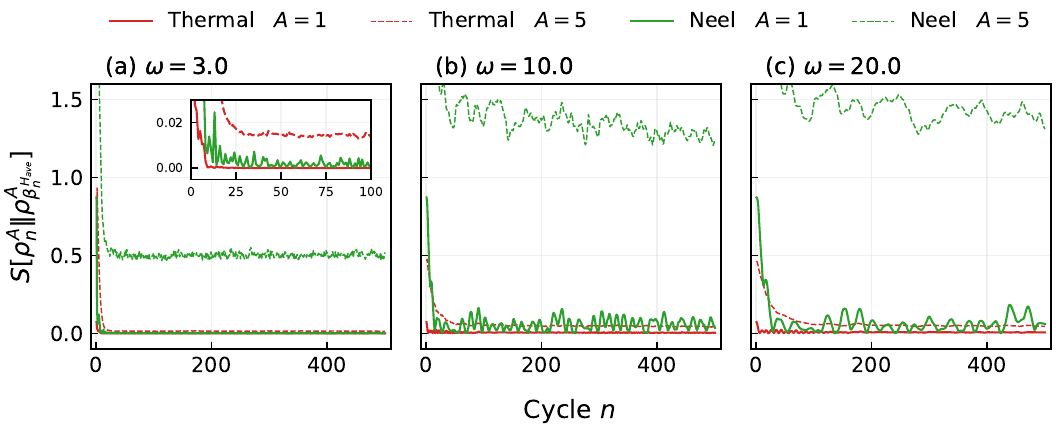}
\caption{\small 
Stroboscopic evolution of relative entropy $S\big[ \rho^A_n \big\Vert  \rho^A_{\beta_n} \big]$ for characterizing the subsystem thermalization of Floquet dynamics for three driving frequencies $\omega=3.0, 10.0, 20.0$ from an initial state which is either a thermal state $\rho_{\beta=1}$ (red) or a Néel state (green) for two different subsystem sizes $A=1$ (solid lines) and $A=5$ (dashed lines). Here, we choose $H_0=H_{\rm ave}$ to determine $\beta_n$ by \eq{be_beta_n}. The inset figure in the subfigure (a) shows the zoom-in pattern in the regime $0 \le n \le 100$. The relative entropies approach nearly-vanishing values steadily in all cases when the initial state is $\rho_{\beta=1}$, implying almost-global thermalization rather than just subsystem thermalization. On the other hand, the relative entropies in cases with the initial Néel state exhibit substantial fluctuations above zero, even for $A=1$, implying no stable subsystem Floquet thermalization.
}
\label{thermal_neel_SA_Have}
\end{figure*} 

\subsubsection{For initial Néel state}

We now adopt the ansatz thermal state \eq{ansatz_thermal} to study Floquet thermalization starting from a Néel state. The results for the stroboscopic evolution of relative entropy $S[ \rho^A_n \big\Vert  \rho^A_{\beta_n}]$ are presented in Fig. \ref{thermal_neel_SA_Have}. For comparison, we also present the results with the initial state being $\rho_{\beta=1}$. 

We first focus on the case where the initial state is the Néel state.
For $A=1$, the relative entropy of the large-$n$ Floquet states approaches zero, although sizable fluctuations persist. 
The amplitude of these fluctuations increases with the driving frequency.
This contrasts with the case of an initial thermal state, where the relative entropy approaches zero with much smaller fluctuations.
For the half-chain subsystem ($A=5$), the relative entropy for the initial Néel state deviates from zero significantly for all three frequencies. {This behavior is consistent with the Néel state remaining globally pure under unitary Floquet evolution, and can only be well approximated by a thermal state locally but not by a global thermal state.} 

In contrast with the large-$n$ relative entropy for the half-chain subsystem ($A=5$) in Fig.~\ref{thermal_neel_SA_Have}, we find that the values obtained from the initial thermal state are substantially smaller than those from the initial Néel state for all three driving frequencies. {It suggests that the quantum coherence of the initial state endures under the unitary Floquet evolutions, and leads to different large-$n$ Floquet states classified by the residual quantum coherence. We will further investigate whether quantum coherence can be destroyed under coarse-graining to yield approximate thermal-state features by studying deviations from the fluctuation theorem. }

\begin{figure*}[bth]
\includegraphics[width= 1.0\textwidth]{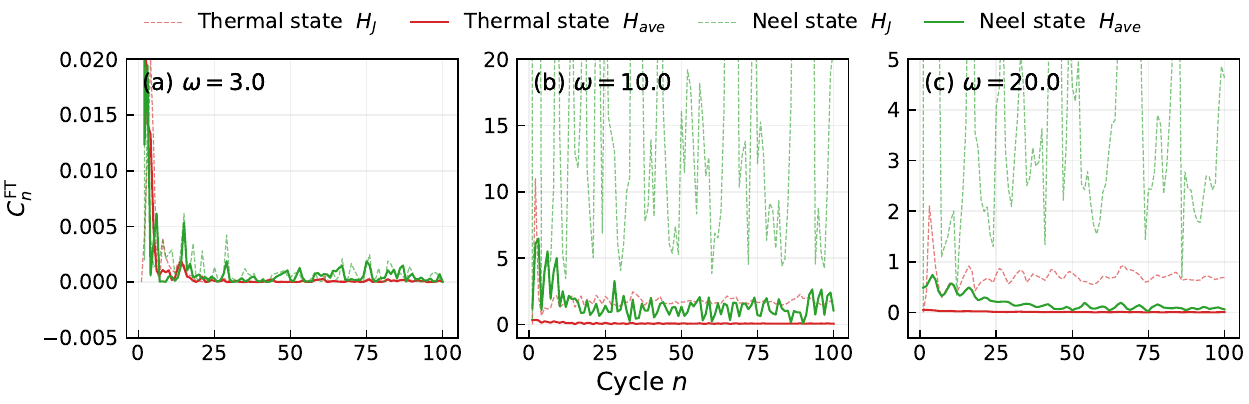}
\caption{\small 
Stroboscopic evolution of $C_n^{\rm FT}$ with $\beta_n$ determined by \eq{be_beta_n} for {choosing $H_{\rm mod} =H_J$ (dashed lines) and $H_{\rm ave}$ (solid lines),} and for the initial state $\rho_0$ which is either a thermal state $\rho_{\beta=1}$ (red) or a Néel state (green). We again consider three driving frequencies (a) $(\omega=3.0)$, (b) $(\omega=10.0)$, and (c) $(\omega=20.0)$. The results imply that the fluctuation theorem associated with $H_{\rm ave}$ and $\rho_0=\rho_{\beta=1}$ for all three driving frequencies is satisfied for large enough $n$ and thus the global thermalization of Floquet dynamics of an initial thermal state. This is consistent with the result in Fig. \ref{thermal_SA_Hall} for subsystem thermalization with $A=5$. On the other hand, the fluctuation theorem cannot be satisfied if the Floquet initial state is the Néel state. It can be attributed to the quantum coherence of the Néel state, which leads to substantial fluctuations in $C_n^{\rm FT}$. Moreover, $H_J$ always yields larger and more fluctuating $C_n^{\rm FT}$ than $H_{\rm ave}$.}
\label{cnFT_vs_n}
\end{figure*} 

\begin{figure}[bth]
\includegraphics[width= 0.45\textwidth]{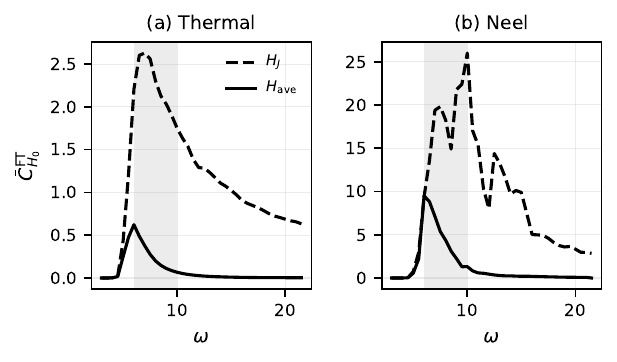}
\caption{\small 
The coarse-graining fluctuation-theorem deviation $\overline{C}^{\rm FT}_{H_0}$ define in \eq{cg_CFT} as a function of the driving frequency $\omega$ for $H_0 = H_J$ (dashed lines) and $H_{\rm ave}$ (solid lines) for the initial state $\rho_0 =\rho_{\beta=1}$ or the Néel state in the subfigures (a) and (b), respectively. 
As shown, $\overline{C}^{\rm FT}_{H_0 = H_{\rm ave}}$ vanishes at both low and high frequencies, but comes with a crossover prethermal bump at intermediate frequencies. 
In contrast, $\overline{C}^{\rm FT}_{H_0 = H_J}$ has a finite value for large $\omega$ and with complicated crossover behaviors for the Néel state. 
}
\label{CFT_average}
\end{figure} 

\subsection{Characterization by fluctuation theorem}

\subsubsection{Stroboscopic evolution of fluctuation-theorem deviation}

An alternative way to characterize global thermalization of the large-$n$ Floquet states is to examine deviations from the fluctuation theorem. We perform the checks by again adopting the ansatz thermal state \eq{ansatz_thermal} to determine $\beta_n$ by \eq{be_beta_n}, and then to evaluate $C_n^{\rm FT}$ of \eq{CnFT_def} for the cases with the initial state to be either $\rho_{\beta=1}$ or the Néel state. The results are shown in  Fig. \ref{cnFT_vs_n}. For comparison, we also show the results for a different ansatz thermal state constructed by using $H_J$.

We first focus on the results by using the ansatz thermal state \eq{ansatz_thermal}. For the case with $\rho_{\beta=1}$ as the initial state, we see that the large-$n$ $C_n^{\rm FT}$ approaches zero for all three frequencies. It implies the global Floquet thermalization. On the other hand, for the case with the Néel state as the initial state, the large-$n$ $C_n^{\rm FT}$ fluctuates significantly above zero for all three frequencies. This observation is consistent with the subsystem thermalization analysis presented above and with the fact that a pure state cannot be globally thermalized under unitary dynamics.

On the other hand, we see the patterns of $C_n^{\rm FT}$ from using the ansatz thermal state associated with $H_J$ are either highly fluctuating for the Néel initial state or significantly deviate from zero for $\rho_{\beta=1}$ initial state. 
It implies that the corresponding ansatz thermal state fails to provide a consistent description of the large-$n$ Floquet states, even though $H_J$ seems a natural choice from the conventional fluctuation theorem of work statistics.
Thus, the success of the fluctuation theorem by choosing the ansatz state \eq{ansatz_thermal} implies that $H_{\rm ave}$ is the modular Hamiltonian of the large-$n$ Floquet states, so that the (modular) fluctuation theorem holds \cite{Pusz:1977hb, Lenard1978ThermodynamicalPO, Benoist:2023tfy, Cirafici:2024ccs}.

\subsubsection{Coarse-graining fluctuatuin-theorem deviation}

The significant fluctuation patterns of $C_n^{\rm FT}$ shown in Fig. \ref{cnFT_vs_n} motivate us to introduce the coarse-graining version of $C_n^{\rm FT}$, such as the following quantity:
{
\be\label{cg_CFT}
\overline{C}^{\rm FT}_{H_0} = {1\over 20} \sum_{n=81}^{100} C^{\rm FT}_n 
\ee
where $H_0=H_{\rm mod}$ can be either $H_J$ or $H_{\rm ave}$, and we have just taken the average of $C^{\rm FT}_n$ over the last 20 Floquet cycles. 
}

Instead of focusing on the detailed stroboscopic evolution, we now consider the coarse-grained fluctuation-theorem deviation and investigate its dependence on the driving frequency over a wide frequency range.
The results are shown in Fig. \ref{CFT_average} for the cases with either $\rho_{\beta=1}$ or the Néel state as the initial state. 
We see that $\overline{C}^{\rm FT}_{H_0}$ approaches zero at low and high frequency regimes if we choose the ansatz thermal state \eq{ansatz_thermal}, and shows a crossover bump in the intermediate frequency regime. This pattern holds for both initial states. 
These results suggest that, after temporal coarse-graining, the large-$n$ Floquet states exhibit nearly thermal behavior in both the low- and high-frequency regimes.

On the other hand, if we use $H_J$ to construct the ansatz thermal state, $\overline{C}^{\rm FT}_{H_0}$ only vanishes in the low frequency regime, and for the case of the initial Néel state, it shows quite complicated patterns in the intermediate frequency regime, not a smooth crossover. This is consistent with the stroboscopic pattern of $C^{\rm FT}_n$ at intermediate and high frequency regimes.

\section{Conclusion}\label{conclude}

We have developed a unified framework for characterizing thermalization in periodically driven quantum many‑body systems by combining subsystem thermalization with work‑statistics diagnostics. Subsystem thermalization provides a local measure of equilibration through the behavior of reduced density matrices, while work statistics quantify global energy absorption and the role of coherent processes through both two‑point measurement distributions and quasi‑probabilistic characteristic functions. Examining these diagnostics within the same Floquet setting reveals that they track the same dynamical crossover as summarized in Fig. \ref{CFT_average}: the emergence of finite-temperature behavior at high-frequency driving regime, then entering a persistent prethermal crossover via gradual loss of coherence by lowering the driving frequency, and the eventual onset of heating toward the infinite‑temperature state that causes dephasing even for an initial pure state.
All these phenomena may be tied to the scrambling of the system's states across different relevant length scales, as dictated by the driving frequency.
This agreement demonstrates that local and global indicators of thermalization are deeply connected and can be used together to resolve the mechanisms governing driven many‑body dynamics. The framework presented here offers a systematic and experimentally accessible approach for quantifying thermalization, coherence decay, and emergent (non-)thermality in periodically driven quantum matter.

 \appendix

\section{Derivation of the Mechanical Average Work}
\label{appA}
In this appendix, we derive the expression for the average work performed by the periodic driving during a Floquet cycle.
For a quantum system with a time-dependent Hamiltonian $H(t)$ and density matrix $\rho(t)$, the average energy is
\be
E(t)= {\rm Tr}\left[H(t)\rho(t) \right].
\ee
Taking the time derivative yields
\be
dE(t) = {\rm Tr}\left[ \frac{\partial H(t)}{\partial t}\rho(t)\right]+{\rm Tr}\left[ H(t)\frac{\partial \rho(t)}{\partial t}\right]
\label{eq_dE}
\ee
The second term describes the energy change due to the evolution of the quantum state. For the Floquet dynamics considered in this work, the system evolves unitarily. Then, using the Schrodinger equation 
\be
{d\rho(t)\over dt}=-i[H(t),\rho(t)]
\ee
Substituting it into Eq.~(\ref{eq_dE}), one finds
\be
&{\rm Tr} \Big( H(t)\frac{\partial \rho(t)}{\partial t} \Big)
=  -i\, {\rm Tr} \Big(  H(t)\big[ H(t),\rho(t) \big ] \Big)
\nonumber\ \\
&= -i\,{\rm Tr} \Big(  H^2(t)\rho(t)-H(t)\rho(t)H(t) \Big)=0 ,
\ee
where the cyclic property of the trace has been used.
Integrating Eq.~(\ref{eq_dE}) over the $n$-th Floquet cycle gives the average work
\be
\langle W \rangle_n &=&
\int_{(n-1)T}^{nT}  {\rm Tr} \left[ \rho(t) \frac{\partial H(t)}{\partial t}\right] dt 
\\
&=& \!E(nT)\!-\!E((n\!-\!1)T\;.
\ee
Since the Hamiltonian at the stroboscopic times is $H_J$, one obtains
\be
\langle W \rangle_n = {\rm Tr}\Big[ H_J(\rho_n-\rho_{n-1}) \Big].
\ee

 \section{Finite-Size Scaling of Coherence Measures}\label{appB}
\begin{figure}[bth]
\includegraphics[width= 0.5\textwidth]{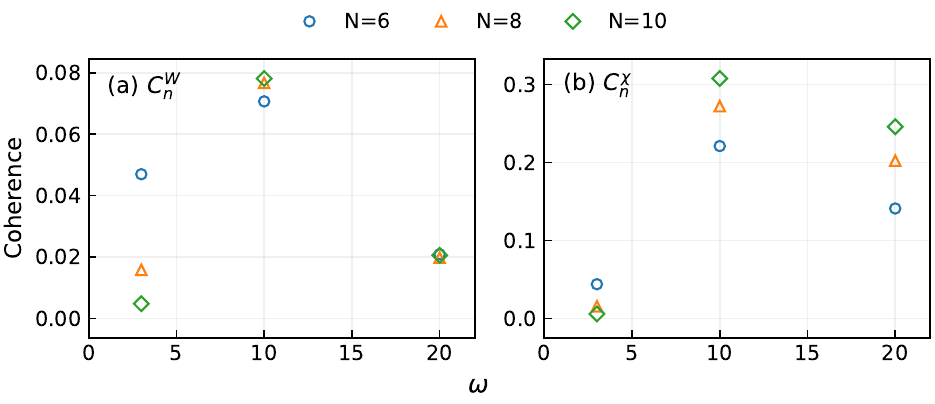}
\caption{\small 
Finite-size scaling analysis of coherence in the periodically driven spin chain. Panel (a) shows the $C_n^W$, while panel (b) displays th $C_n^{\chi}$, for system sizes $N=6, 8$, and $10$. Both quantities are plotted as functions of the driving frequency $\omega$. The weak size dependence near the coherence peak suggests that the observed behavior persists in larger systems.
}
\label{coherence_FSS}
\end{figure}

To examine whether the observed coherence behavior is affected by finite-size effects, we perform a finite-size scaling analysis for system sizes $N=6$, $8$, and  $10$. The results are summarized in Fig.~\ref{coherence_FSS}.

Figure~\ref{coherence_FSS}(a) shows the coherent  $C_n^{W}$, while Fig.~\ref{coherence_FSS}(b) presents the coherence $C^{\chi}_n$, both evaluated after the Floquet evolution at different driving frequencies $\omega$. For all system sizes, the coherence exhibits a non-monotonic dependence on the driving frequency. In particular, both  $C_n^{W}$ and $C^{\chi}_n$ are strongly suppressed in the high-frequency regime, increase substantially at intermediate frequencies, and decrease again as the driving frequency is further reduced.

Importantly, the overall frequency dependence remains qualitatively unchanged as the system size increases. The positions of the coherence maxima are nearly size-independent, and the magnitude of the coherence varies only moderately between different system sizes. In particular, the enhancement of coherence around the intermediate-frequency regime persists from $N=6$ to $N=10$, indicating that this feature is not solely a consequence of finite-size effects.

The finite-size scaling analysis demonstrates that the qualitative behavior of both the coherent $C_n^{W}$ and the$C^{\chi}_n$ survives as the system size increases, providing evidence that the coherence generation mechanism remains relevant in larger systems and potentially persists in the thermodynamic limit.

\bibliography{ref}

\end{document}